\documentclass[epj]{svjour}
\usepackage{amsmath,amssymb}
\usepackage{tensor}
\usepackage{etoolbox}
\usepackage{graphicx}
\usepackage{textgreek}
\usepackage{microtype}
\usepackage{bm}
\usepackage{multirow}

\usepackage[utf8]{inputenc}
\usepackage[ngerman,english]{babel}
\usepackage{mathtools}

\usepackage{bm}
\usepackage{epstopdf}
\usepackage{amsbsy}
\usepackage{amsmath}
\usepackage{amssymb}
\usepackage[separate-uncertainty = true,multi-part-units=single]{siunitx}
\usepackage{commath}
\usepackage{cite}
\DeclareSIUnit{\fm}{\femto\metre}
\newcommand{\sNN}{\sqrt{s_{NN}}}

\newcommand*\diff{\mathop{}\!{d}}
\newcommand*\Diff[1]{\mathop{}\!{d^#1}}
\newcommand{\ybeam}{y_\mathrm{beam}}
\newcommand{\EQ}[3]{
  \begin{equation}
    \label{#1}
    #2
    \;#3
  \end{equation}
}

\begin{document}
\title{Centrality dependence of limiting
fragmentation}
\author{B.\,Kellers \and G.\,Wolschin}
\institute{Institut f\"ur Theoretische Physik der Universit\"at Heidelberg, Philosophenweg 12-16, D-69120 Heidelberg, Germany, EU}
\date{Received: date / Revised version: date} 
\abstract{
We investigate the centrality-dependent validity of the limiting-fragmentation hypothesis in relativistic heavy-ion collisions at energies reached at the Relativistic Heavy Ion Collider (RHIC) and the Large Hadron Collider (LHC).
A phenomenological analysis of  Au-Au and Pb-Pb collisions
within a three-source relativistic diffusion model (RDM) is performed at $\sNN=19.6, 62.4, 130, 200, 2760$ and $5023$ GeV using four centrality cuts at each energy. Linear and nonlinear expressions for the rapidity drift function are tested and their physical relevance is discussed. Our results are compatible with the limiting-fragmentation conjecture for the investigated centralities in the full energy range. The number of particles in the fragmentation and fireball sources are found to depend on $\sqrt{s_{NN}}$  logarithmically and cubic-logarithmically, respectively. 
\PACS{
	{25.75.-q}{Relativistic heavy-ion collisions}
	\and
	{24.10.Jv}{Relativistic models (nuclear reactions)}
	}
}
\maketitle
\section{Introduction}

The occurence of limiting fragmentation (LF), or extended longitudinal scaling, had been predicted for hadron-hadron and electron-proton collisions by Benecke et al. \cite{benecke69}. It was first shown to be present in charged-hadron production at large pseudorapidities in the fragmentation region of
$p\bar{p}$ data, in an energy range of $\sqrt{s}=53$--$900$\,GeV \cite{al86}: The charged-particle pseudorapidity yield $dN/d\eta$ does not depend on energy over a large range of pseudorapidities $\tilde{\eta}=\eta - y_\text{beam}$, with the beam rapidity $y_\text{beam}$.
The fragmentation region grows in pseudorapidity with increasing collision energy and can cover more than half of the pseudorapidity range over which particle production occurs. The approach to a universal limiting curve is a characteristic feature of the particle production process, which turns out to be especially outstanding in relativistic heavy-ion collisions.

At the Relativistic Heavy Ion Collider RHIC in Brook-haven, limiting fragmentation was shown to occur in Au-Au collisions in the energy range $\sNN$ = \SIrange{19.6}{200}{\GeV} \cite{bea02,bb03,ada06}. For a given centrality, the pseudorapidity distributions 
of produced charged particles were found to scale with energy according to the LF hypothesis in a given centrality class. 

It is presently an open question whether limiting fragmentation will persist at the much higher incident energies of $\sNN$ = 2.76 and 5.02 TeV that are available at the CERN Large Hadron Collider (LHC) in Pb-Pb collisions since experimental results in the fragmentation region are not available due to the lack of a dedicated forward spectrometer. Nevertheless, it is most interesting to account for the collision dynamics more completely in this region. In Ref.\,\cite{kgw19}, we had therefore studied central Pb-Pb collisions at LHC energies in a phenomenological model, with the result that LF can be expected to hold in central events. 

We now extend this work to investigate the centrality dependence of limiting fragmentation in a three-source relativistic diffusion model (RDM, \cite{gw15}). We consider equivalent centrality classes at four RHIC energies and two LHC energies in the range $0-30$\,\%. At RHIC energies, a detailed comparison with PHOBOS data \cite{alver11} is possible in all four centrality bins, whereas at LHC energies our analysis remains a model-dependent prediction. For all six energies and four centralities, we also deduce the number of produced charged hadrons in the respective two fragmentation sources and the fireball particle-production source, and determine their dependencies on 
$\sqrt{s_{NN}}$.

Our analysis complements microscopic approaches such as the multiphase transport model AMPT by Ko et al.\;\cite{ko05} or HIJING \cite{wang91,papp18} in order to assess whether centrality-dependent LF is valid from RHIC to LHC energies.  AMPT  \cite{ko05} had been tuned for the most central bin at RHIC and LHC energies \cite{basu16}. In spite of disagreements with the LHC data in the midrapidity region,  it had been concluded \cite{na11} that AMPT and other microscopic codes reproduce LF at RHIC energies. The same conclusion had been drawn from calculations in the color-glass-condensate framework \cite{sta06}. LF as an initial-state probe has been investigated for several models in 
Ref.\,\cite{tor19}.

There exist also other phenomenological approaches such as the thermal model \cite{hag65,pbm95,mabe08,pbm16} and hydrodynamical models \cite{hesne13} which offer predictions regarding LF. The thermal model is appropriate to calculate particle production rates near midrapidity, but cannot be expected to predict distribution functions at forward rapidities that are needed to check the LF conjecture. It has nevertheless been employed in the forward region \cite{cleymans08}, concluding that LF should be violated at LHC energies. A recent phenomenological study \cite{basu20} that uses various fit functions to extrapolate to the unmeasured forward $\eta$-region finds instead that LF should hold at the TeV energy scale.

In the string percolation model \cite{bro07}, a delicate energy-momentum compensation is required for LF to be exact, and the authors show that this is unlikely to occur. Inspite of this, their model yields reasonable agreement with the RHIC Au-Au data in the fragmentation region and it appears that the percolation model is consistent with
 LF in an approximate sense even for Pb-Pb at 5.5 TeV. The consequences of a violation of exact LF were further investigated in Refs.\,\cite{deus08,bau12}. 
 
The relativistic diffusion model is based on three sources for particle production: a midrapidity fireball source and two fragmentation sources \cite{wolschin99,biya02,wobi06,wob06,wolschin13,gw16}.
The time evolution of the distribution functions is accounted for through solutions of a Fokker-Planck equation (FPE) for the rapidity variable which are subsequently transformed to pseudorapidity space through the appropriate Jacobian. The three sources can then be added to obtain the charged-hadron distribution
that is used to check the LF hypothesis.

In Ref.\,\cite{hgw20} we have shown that a  FPE in rapidity space can be derived from a nonequilibrium-statistical theory of non-Markovian processes in spacetime that are equivalent \cite{deb88,dunkel05} to relativistic Markov processes in phase-space. The fluctuating background that is required for such a description to be valid is provided by the quarks and gluons in the fragmentation sources and in the fireball. A thermally equilibrated heat bath as in the theory of Brownian motion is not needed in the derivation. 
One obtains a FPE for time-dependent particle transport in rapidity space. Drift and diffusion terms are related through a fluctuation-dissipation relation (FDR). With a constant diffusion coefficient and the FDR,  the drift function in stopping can be calculated from the condition that the stationary solution of the FPE equals a distribution function derived in the color-glass condensate framework. For particle production, a similar path will be pursued.

In the present work about charged-hadron production and limiting fragmentation, we also use a FPE with constant diffusion coefficients, and either a linear dependence of the drift on rapidity $y$ as in the original phenomenological RDM, or a $\sinh(y)$ dependence \cite{fgw17,kgw19} that asymptotically leads to the Maxwell-J\"uttner equilibrium distribution. In case of the nonlinear sinh-drift, the transport equation itself still remains linear, such that the time dependence of the three sources can be treated individually, and the corresponding results be added incoherently.


We briefly summarise the basic formulation in the next section. In sect.\,3, we apply the model with sinh-drift term that requires a numerical solution to calculate charged-hadron pseudorapidity distributions in Au-Au and Pb-Pb at RHIC and LHC energies in four different centrality bins. In each case, the LF conjecture is tested. For central collisions, results are compared with analytical solutions obtained earlier for linear drift. In sect.\,4, we determine the number of produced charged hadrons in the fragmentation sources and the fireball source at all six energies and four centralities, and discuss their energy dependence. The conclusions are drawn in sect.\,5.

\section{A phenomenological three-source model}
\label{sec2}
The Lorentz-invariant cross section for produced particles in relativistic heavy-ion collisions is
\EQ{eq:3e}{E \frac{\Diff3 N}{\diff^3 p} = \frac{\Diff2 N}{2 \pi p_\text{T} \diff p_\text{T} \diff y} = \frac{\Diff2 N}{2 \pi m_\text{T} \diff m_\text{T} \diff y}}{}
with the energy $E = m_\text{T}\cosh(y)$, the transverse momentum $p_\text{T} = \sqrt{p_{x}^2+p_{y}^2}$, the transverse mass $m_\text{T} = \sqrt{m^2+p_\text{T}^2}$,  and 
the rapidity $y=0.5\ln[(E+p_{\parallel})/(E-p_{\parallel})]$. 

In a three-source model for particle production, 
the rapidity distributions for all three sources $k=1,2,3$ are obtained by integrating over the transverse mass $m_\text{T}$
\EQ{rap}{
  \frac{\diff N_k}{\diff y}(y,t) = c_{k}^{\,b} \int\; m_\text{T} E \frac{\Diff3 N_k}{\diff^3 p} \diff m_\text{T}
}{,}
where the normalisation constants $c_k^{\,b}$ for the three sources depend on centrality, or impact parameter $b$. (Here and subsequently we omit the index $b$ in all other variables such as $N_k$). The experimentally observable distribution {$dN/dy$ is evaluated in every centrality bin at the freeze-out time, $t = \tau_\text{\,f}$\,, which corresponds to the interaction time $\tau_\text{int}$ of Refs.\,\cite{wolschin99,fgw17}: the time during which the system interacts strongly. The full rapidity distribution function for produced charged hadrons is obtained by weighting the three partial distribution functions that have been integrated over the transverse masses with the respective numbers of produced charged hadrons $N_\text{ch}^{k}$ and adding them  according to
\begin{eqnarray}
\frac{dN_\text{ch}}{dy}(y,t=\tau_\text{f})=\frac{dN_{1}}{dy}(y,\tau_\text{f})
 +\frac{dN_{2}}{dy}(y,\tau_\text{f})\\\nonumber
+\frac{dN_\text{gg}}{dy}(y,\tau_\text{f})\,,
\label{normloc1}
\end{eqnarray}
where the subscript gg refers to gluon-gluon collisions as being the main source in the fireball. Each of the three partial distributions is a product of the number of produced particles in the respective source and the corresponding distribution function $R_k(y,t=\tau_\text{f})$
\begin{equation}
\frac{dN_{k}}{dy}(y,\tau_\text{f})=N_kR_k(y,\tau_\text{f})\,.
\end{equation}
Within the model with sinh-drift that we consider in this work, the superposition of particles from the three sources is still possible because the FPE that governs the time-dependence of the distribution functions is a linear partial differential equation. For symmetric systems, the problem is simplified by only considering the solution for the positive rapidity region and mirroring the result at $y<0$.
\begin{figure}[t!]
\begin{center}
\includegraphics[scale=0.86]{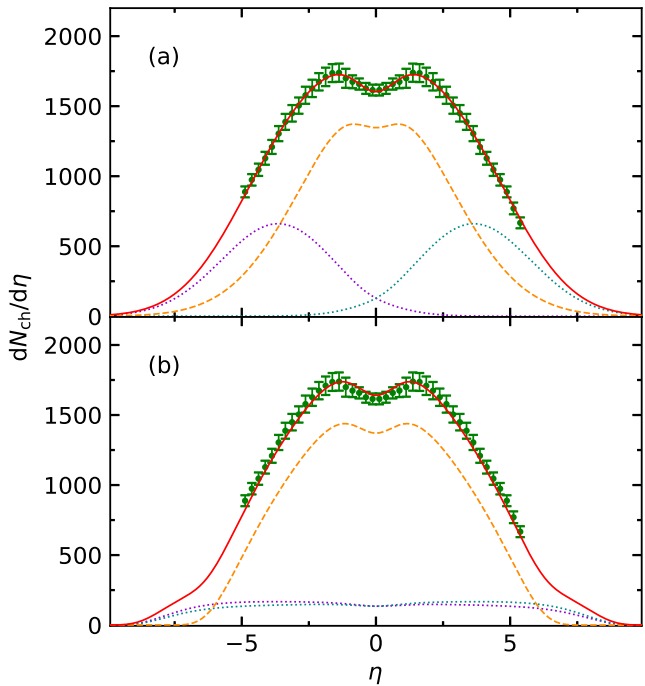}
\caption{(a) Analytical solution of the RDM with linear drift in a $\chi^2$-minimization with respect to the most central \SI{2.76}{TeV} Pb-Pb ALICE data\,\cite{abb13} for charged-hadron production as in Refs.\,\cite{gw15,kgw19}. The solid curve is the overall distribution,  the incoherent sum of the three sources. The distributions resulting from the fragments are symmetric (dotted). The fireball source is the essential contribution to the charged-hadron yield (dashed). (b) {Numerical solution of the RDM with nonlinear drift. The distributions resulting from the fragments cover the full pseudorapidity range (dotted). The midrapidity source is wider and more pronounced compared to the linear-drift case (dashed).}}
\label{fig1}
\end{center}
\end{figure}
The parameters of the three-source model will be determined via $\chi^2$-minimization with respect to the available data, and can be used in extrapolations and predictions.

In view of the high temperatures reached in relativistic collisions at RHIC and LHC energies, we rely on Boltzmann-Gibbs statistics and adopt the Maxwell-J{\"u}ttner distribution
as the thermodynamic equilibrium distribution for $t\rightarrow \infty$ 
\begin{eqnarray}
\label{equ}
  E \frac{d^3N}{d^3p} \Bigr|_\text{eq}\propto E \exp\left(-E/T\right)\\\nonumber
= m_\text{T} \cosh\left(y\right) \exp\left(-m_\text{T} \cosh(y) / T\right).
\end{eqnarray}
 In the relativistic diffusion model \cite{wolschin99,biya02,wolschin13,hgw20,fgw17}, the partial distribution functions $R_k(y,t)$ $(k=1, 2, \text{gg})$ evolve in time towards this thermodynamic equilibrium distribution  through solutions of the Fokker-Planck equation 
\begin{eqnarray}
  \pd{}{t}R_k(y,t) =-\pd{}{y}\left[J_k(y,t)R_k(y,t)\right] \\\nonumber
  + \pd[2]{}{y}\left[D_k(y,t)R_k(y,t)\right]\,,
\end{eqnarray}
with drift functions $J_k(y,t)$ and diffusion functions $D_k(y,t)$. The latter are taken to be constant coefficients $D_k$ in this work. If, in addition,
the drift functions are assumed to be linearly dependent on the rapidity variable $y$, 
the FPE has the Ornstein-Uhlenbeck form \cite{uo} and can be solved analytically \cite{wolschin99}. For $t\rightarrow \infty$ all three subdistributions approach a single Gaussian in rapidity space which is centered at midrapidity $y=0$ for symmetric systems, or at the appropriate equilibrium value $y=y_\text{eq}$ for asymmetric systems.
The equilibrium distribution deviates, however, slightly from the Maxwell-J{\"u}ttner distribution, although the discrepancies are small and become visible only for sufficiently large times.  

In order to attain the correct stationary solution, the drift terms must be modified according to \cite{lavagno02,fgw17}
\begin{align}
J_k(y,t)&=-A_k \sinh(y)\,,
\end{align}
with drift amplitudes
\begin{equation}
A_k= m^k_\text{T} D_k/T\,.
\label{fdt}
\end{equation}
The drift forces in the fragmentation sources $k = 1,2$ grow with increasing distance in $y$-space from midrapidity:
They are maximal at $y=y_\text{beam}$ and minimal at $y=0$ for symmetric systems. The drift is compatible with the tendency of the system to maintain overall energy-momentum conservation. Its occurence in particle production corresponds to the slowing down of the fragmentation sources in the course of the collision. 

The rapidity distribution at thermal equilibrium -- when the three subdistributions would have merged -- can then be derived \cite{fgw17} using Eqs.\,(\ref{rap}) and (\ref{equ}) as
\begin{eqnarray}
\frac{dN_\text{eq}}{dy}=C_{b} \left( m_\text{T}^2 T + \frac{2 m_\text{T} T^2}{\cosh y} + \frac{2 T^3}{\cosh^2 y} \right) \nonumber\\
\times\exp\left({-\frac{m_\text{T} \cosh y}{T}}\right),
\label{eqfdt}
\end{eqnarray}
with $C_{b}$ being proportional to the overall number of produced charged hadrons  $N^\text{tot}_\text{ch}$ in the respective centrality bin. The actual distribution functions remain far from thermal equilibrium and the total particle number is evaluated based on the nonequilibrium solutions of the FPE, which are adjusted to the data in $\chi^2$-minimizations. 

We determine
the drift amplitudes $A_k$ in each centrality bin from the position of the fragmentation peaks. Diffusion coefficients can then be calculated as $D_k=A_k T/m_\text{T}$ from eq.\,(\ref{fdt}). These refer, however, only to the diffusive processes. Since the fireball source and both fragmentation sources also expand collectively, the actual distribution functions are much broader \cite{gw15} than what is obtained from eq.\,(\ref{fdt}). We therefore use values
for the diffusion coefficients (or the widths of the partial distributions) that are adapted to the data. The total particle number is then obtained in each centrality bin from the integral of the overall distribution function. We shall investigate four centrality bins at both, RHIC and LHC energies.

The RDM with linear drift has analytical solutions that can be used directly in $\chi^2$-minimizations with respect to the data, but numerical solutions of the FPE are required for the sinh-drift, as outlined in Ref.\,\cite{fgw17}.
To arrive at a usable form for the computer{,} we transform the equations for $R_k(y,t)$ into dimensionless equations through the introduction of a timescale $t_c$ that defines the dimensionless time variable $\tau = t / t_c$, such that
 $\pd{}{t} = \pd{}{\tau} t_c^{-1}$ and 
\EQ{eq:4}{
  \pd{R_k}{\tau}(y,\tau) = t_cA_k\pd{}{y}\left[\sinh(y)R_k(y,\tau)\right] + t_cD_k\pd[2]{R_k}{y}(y,\tau)
}{.}
With {$A_k = m^k_\text{T} D_k/T$}, we set {$t_c =T/(m^k_\text{T} D_k) = A_k^{-1}$}.
As a result, the dimensionless eqs.\,(\ref{eq:4aa}) depend only on the ratios {$\gamma_k = T/m^k_\text{T}$} of temperature $T$ and transverse masses $m^k_\text{T}$ which measure the diffusion strengths,
\EQ{eq:4aa}{
    \pd{R_k}{\tau}(y,\tau) = \pd{}{y}\left[\sinh(y)\;R_k(y,\tau)\right] + \gamma_k\;\pd[2]{R_k}{y}(y,\tau)
}
{.}
To recover the drift and diffusion coefficients, one therefore has to specify a time scale.
Since the drift determines the peak position{,} we choose the time-like variable $\tau$ initially such that the peak position of the experimental data is reproduced. It is then recalibrated
through the $\chi^2$-minimization routine together with
 the diffusion strengths $\gamma_{k}$ as free parameters. Hence, the three partial distributions in each centrality bin are determined through the parameters $\tau$ and $\gamma_k$, with the two values 
 $\gamma_{1,2}$ for the fragmentation sources being identical for symmetric systems such as Pb-Pb, but differing for asymmetric systems like $p$-Pb.

The numerical  solution is obtained using \textsc{matlab}'s integration routine \texttt{pdepe} for solving parabolic-elliptic partial differential equations.
We had shown in Ref.\,\cite{fgw17} that this method is very accurate when compared to results of finite-element methods such as DUNE \cite{ba10} and FEniCS \cite{aln15}. 
\section{Charged-hadron production and limiting fragmentation}
\label{sec3}
We insert physical values for the temperature $T$, the transverse mass $m_\text{T}$, and the initial conditions $f_\text{i}(y,t=0)$ in order to compare the model results to centrality-dependent data. 
Two distributions centered at the beam rapidities  $y_\text{beam} =\pm \ln(\sqrt{s_{NN}}/m_p)$ with a small width that corresponds to the Fermi motion represent the incoming ions before the collision.
The exact width of the initial distribution does not have a large effect on the time evolution \cite{fgw17}, we use gaussians with $\sigma=0.1$. 
The same standard deviation is assumed for the initial condition of
the midrapidity source, which is centered at $y=0$ for a symmetric system, and at  $y=y_\text{eq}$ for asymmetric systems. 

For the temperature{,} the critical value $T=T_\text{cr}=\SI{160}{\MeV}$ of the cross-over transition between hadronic matter and quark-gluon plasma is adopted. Experimental values are deduced for the transverse mass
 from measured transverse-momentum distributions.

 
We solve eq.\,(\ref{eq:4aa}) numerically for each centrality class and transform the results to rapidity distributions as discussed in Ref.\,\cite{kgw19} according to
\EQ{rapf}
{
  \frac{dN_k}{dy}(y,\tau) = C_b \int m^2_\text{T} R_k(y,\tau)\, d m_\text{T} \\\,.
}{}
The constant ${C_{b}}$ is adjusted to the total number of produced charged hadrons in a given centrality bin $b$.
 \begin{figure}[t!]
\begin{center}
\includegraphics[scale=0.86]{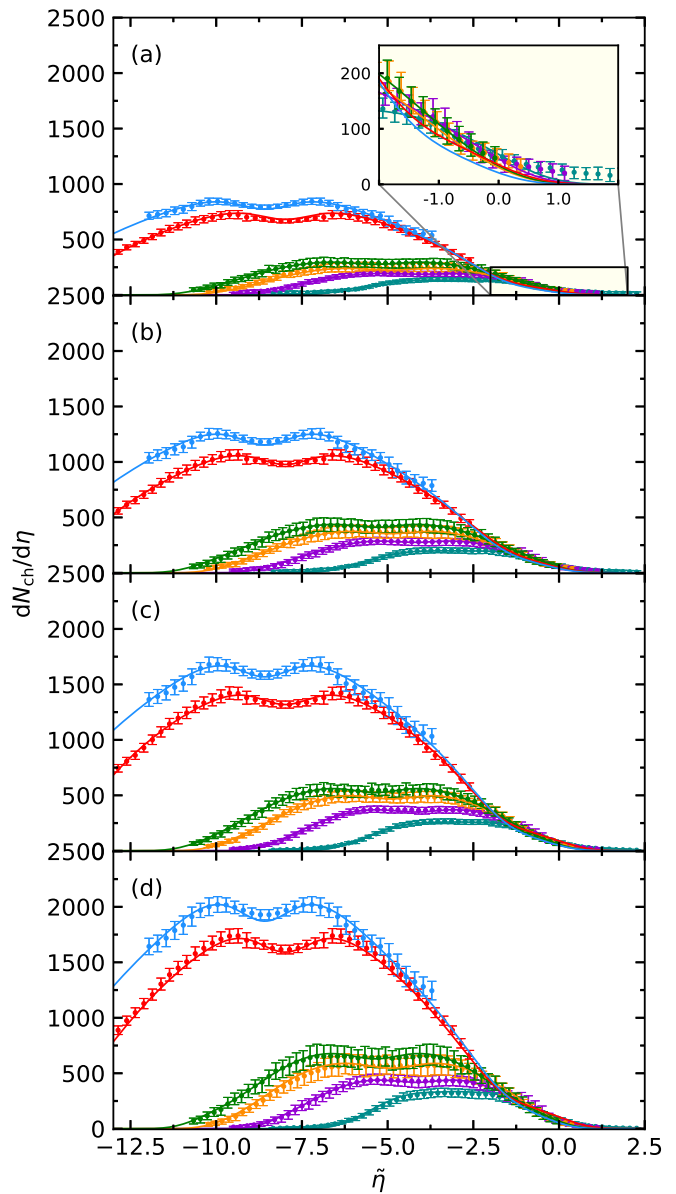}
\caption{Numerical RDM-solutions with nonlinear drift (solid lines) adapted in $\chi^2$-fits to the data (dots) in four centrality classes. From top to bottom in each panel: \SI{5.02}\,{TeV} and \SI{2.76}\,{TeV} Pb-Pb \cite{abb13,alice17}; \SI{200}\,{GeV}, \SI{130}\,{GeV}, \SI{62.4}\,{GeV}, and \SI{19.6}\,{GeV} Au-Au \cite{alver11}. (a) Centralities $20-30$\,\%  at RHIC and LHC. (b) $10-20$\,\%  at RHIC and LHC. (c) $6-10$\,\% at RHIC and  $5-10$\,\% at LHC. (d)  $0-6$\,\% at RHIC and $0-5$\,\% at LHC. 
}
\label{fig2}
\end{center}
\end{figure}
 \begin{figure}
\begin{center}
\includegraphics[scale=0.86]{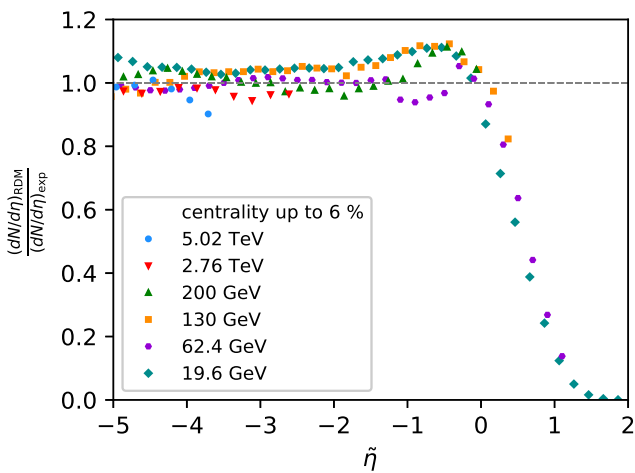}
\caption{Ratio plot of numerical RDM-solutions to data for central collisions as in fig.\,\ref{fig2}\,(d), $0-6$\,\% at RHIC and $0-5$\,\% at LHC. The ratio ${(dN/d\eta)_\mathrm{RDM}}/{(dN/d\eta)_\mathrm{exp}}$ is reasonably close to unity in the range $\tilde{\eta}<0$ for LF to be approximately fulfilled.
For $\tilde{\eta}>0$ the ratio approaches zero rapidly, since $(dN/d\eta)_\mathrm{RDM}$ drops towards zero faster than the data to fulfil the Dirichlet boundary condition imposed by the numerical solver of the FPE.
}
\label{fig2x}
\end{center}
\end{figure}
\begin{table*}[h]
\caption{System and model parameters  with sinh-drift in four centrality classes for Au-Au (RHIC) and Pb-Pb (LHC) at six incident energies. Listed are particle content $N_\text{gg}$ and $N_{1,2}$  of the fireball and fragmentation sources, corresponding diffusion strengths $\gamma_{\text{gg}}$ and $\gamma_{1,2}$, time-like variable $\tau$ (see text),  $\chi^2$- and $\chi^2/\text{ndf}$-values.}
\centering
\begin{tabular}{rrrrrrrrr}
\hline
\hline
\multicolumn{1}{c}{$\sNN$ (GeV)} & 
\multicolumn{1}{c}{$|\ybeam|$} & 
\multicolumn{1}{c}{$N_\mathrm{gg}$} & 
\multicolumn{1}{c}{$N_\mathrm{1,2}$} & 
\multicolumn{1}{c}{$\gamma_\mathrm{gg}$} & 
\multicolumn{1}{c}{$\gamma_\mathrm{1,2}$} & 
\multicolumn{1}{c}{$\tau$} &
\multicolumn{1}{c}{$\chi^2$} &
\multicolumn{1}{c}{$\chi^2/\mathrm{ndf}$} \\
\hline

\multicolumn{9}{c}{Centrality: 0--6\,\% (RHIC) and 0--5\,\% (LHC)}\\ 
\hline

19.6 & 3.037& 20 & 830 & 0.214 & 6.01 & 0.800 & 239.1 & 4.78\\

62.4 & 4.197 & 435 & 1273 & 4.85 & 18.8 & 0.706 & 80.1 & 1.60\\
130 & 4.931 & 1276 & 1533 & 11.6 & 43.3 & 0.681 & 60.2 & 1.20\\

200 & 5.362 & 2013 & 1656 & 11.0 & 61.1 & 0.578 & 23.9 & 0.48\\
2760 & 7.987 & 12638 & 2309 & 115 & 1333 & 0.050 & 28.9 & 0.76\\

5023 & 8.585 & 16196 & 2548 & 232 & 1424 & 0.027 & 32.2 & 1.07\\
\hline
 
\multicolumn{9}{c}{Centrality: 6--10\,\% (RHIC) and 5--10\,\% (LHC)} \\
\hline

19.6 & 3.037 & 20 & 700 & 0.274 & 6.66 & 0.800 & 237.7 & 4.75\\

62.4 & 4.197 & 439 & 1068 & 6.69 & 20.4 & 0.556 & 93.9 & 1.88\\
130 & 4.931 & 1228 & 1271 & 10.5 & 42.4 & 0.337 & 18.0 & 0.36\\

200 & 5.362 & 1871 & 1310 & 16.9 & 63.7 & 0.222 & 14.9 & 0.30\\
2760 & 7.987 & 10380 & 1987 & 117 & 1324 & 0.054 & 9.5 & 0.25\\

5023 & 8.585 & 13814 & 1926 & 237 & 1612 & 0.029 & 19.7 & 0.66\\
\hline

\multicolumn{9}{c}{Centrality: 10--20\,\% (RHIC and LHC)} \\
\hline

19.6 & 3.037 & 17 & 560 & 0.356 & 7.74 & 0.748 & 259.4 & 5.19\\

62.4 & 4.197 & 331 & 844 & 4.13 & 24.7 & 0.628 & 73.2 & 1.46\\
130 & 4.931 & 899 & 977 & 11.1 & 42.8 & 0.280 & 24.8 & 0.50\\

200& 5.362 & 1403 & 1081 & 19.0 & 68.5 & 0.175 & 25.5 & 0.51\\
2760& 7.987 & 7730 & 1573 & 141 & 955 & 0.043 & 5.66 & 0.15\\

5023 & 8.585 & 10153 & 1597 & 225 & 1565 & 0.030 & 18.1 & 0.60\\
\hline
 
\multicolumn{9}{c}{Centrality: 20--30\,\% (RHIC and LHC)} \\
\hline

19.6 & 3.037 & 18 & 391 & 0.539 & 9.22 & 0.792 & 277.1 & 5.54\\

62.4 & 4.197 & 273 & 553 & 6.21 & 25.0 & 0.297 & 78.3 & 1.56 \\
130 & 4.931 & 639 & 598 & 11.9 & 42.2 & 0.404 & 93.5 & 1.87\\

200& 5.362 & 1013 & 697 & 19.6 & 66.2 & 0.198 & 9.51 & 0.19 \\
2760& 7.987 & 5520 & 982 & 122 & 1073 & 0.050 & 13.5 & 0.36\\

5023 & 8.585 & 7027 & 959 & 227 & 1458 & 0.030 & 12.6 & 0.42\\
\hline
\hline
\end{tabular}
\label{tab1}
\end{table*}

At LHC energies, the fireball source yields the largest contribution to charged-hadron production. Particles that are produced from the fragmentation sources are not distinguishable from those originating from the fireball, but must be included in a phenomenological model. In particular, when regarding the limiting-fragmentation conjecture, the role of the fragmentation distributions is decisive since they determine the behavior of the distribution functions at large values of rapidity. 

In case of unidentified charged particles, we first have to transform from rapidity- to pseudorapidity space in order to directly compare to data. The  scattering angle $\theta$ determines the pseudorapidity variable $\eta$ according to
\begin{equation}
\eta=\frac{1}{2}\ln\frac{|\bf{p}|+\rm p_\parallel}{|\bf{p}|-\rm p_\parallel}=-\ln\left[\tan\left(\theta/2\right)\right]\,,
\label{eta}
\end{equation}
and we obtain the pseudorapidity distribution function ${{d}N}/{{d}\eta}$ from the rapidity distribution ${{d}N}/{{d}y}$ as
\begin{align}
	\frac{{d}N}{{d}\eta} \simeq \frac{{d}y}{{d}\eta} \frac{{d}N}{{d}y} = \mathcal{J}\left(\eta, \frac{m}{p_\text{T}}\right) \,\frac{{d}N}{{d}y}\,,
	\label{dNdeta}
\end{align}
with the Jacobian
\begin{align}
\mathcal{J}\left(\eta, \frac{m}{p_\text{T}}\right) = 
		\left[{{ 1+\left(\frac{m}{p_\text{T}\cosh(\eta)}\right)^2}}\right ]^{-1/2}
	\label{jac}
\end{align}
for produced particles with mass $m$ and transverse momentum $p_\text{T}$.
Since the transformation depends on the squared ratio $(m$/$p)^2$ of mass and momentum $p=p_\text{T}\cosh(\eta)$ of the produced particles, its effect increases with the mass of the particles and is most pronounced at small momenta. The full $p_\text{T}$-distributions are, however, not available for all particle species that are included in the pseudorapidity measurements and hence, one has to make estimates.

 We had determined in Ref.\,\cite{wolschin12} the Jacobian $\mathcal{J}_0$ at $\eta=y=0$ in central 2.76 TeV Pb-Pb collisions for identified $\pi^-, K^-$, and antiprotons from the experimental values $\frac{{d}N}{{d}\eta}|_\text{exp}$ and $\frac{{d}N}{{d}y}|_\text{exp}$ as $\mathcal{J}_0=0.856$. Values at the other energies and centralities were found to vary between $\mathcal{J}_0=0.830$ and 0.861. With eq.\,(\ref{jac}) for $p_\text{T}\equiv \langle p_\text{T}^{\text{eff}}\rangle$ one obtains 
\begin{equation}
\langle p_\text{T}^{\text{eff}}\rangle=\frac{\langle m \rangle \mathcal{J}_0}{\sqrt{1-\mathcal{J}_0^2}}\,.
\label{pteff}
\end{equation}
The mean mass $\langle m \rangle$  is calculated from the abundancies of pions, protons and kaons. Using $\mathcal{J}_0$,
the Jacobian can be written independently from the values of $\langle m \rangle$ and $\langle p_\text{T}^{\text{eff}}\rangle$ as
\begin{equation}
\mathcal{J} \left(\eta, \mathcal{J}_0\right) = 
	\left[1+\left(\frac{\sqrt{1-\mathcal{J}_0^2}}{\mathcal{J}_0\cosh(\eta)}\right)^2\right]^{-1/2}\,,
	\label{jac1}
\end{equation}
 resulting in $\mathcal{J}(\eta)=[1+0.365/\cosh(\eta)^2]^{-1/2}$ for central 2.76 TeV Pb-Pb collisions, and the same value at 5.02 TeV.

The effect of the Jacobian is most pronounced near midrapidity, where it generates the dip in the pseudorapidity distributions, as can be seen in fig.\,\ref{fig1}. Here, calculations in the RDM using both, linear drift \cite{gw16} and sinh-drift \cite{kgw19}, are compared with ALICE data for central Pb-Pb at 2.76 TeV \cite{abb13}. The parameters and $\chi^2$-values for the sinh-drift are included in table\,\ref{tab1}, the linear-drift calculation is as in Ref.\,\cite{gw16}. 

We find that the values of the diffusion strengths $\gamma_{1,2}$ in the fragmentation sources and $\gamma_\text{gg}$ in the fireball source that are needed to fit the data are consistently larger than what is calculated from the relation $\gamma_k=T/m^k_\text{T}$ in the sinh-drift model. 
The underlying reason is the additional collective expansion in the three sources, as had been discussed before in more detail for the linear-drift case \cite{wolschin13,gw16} and in the sinh-drift model \cite{kgw19}.

Whereas the numbers of produced particles in the fireball as well as in both fragmentations sources have strong monotonic dependences on energy and on centrality, this is not the case for the diffusion strengths $\gamma_{gg}$ and $\gamma_{1,2}$, see table\,\ref{tab1}. These coefficients as determined from $\chi^2$-minimizations rise monotonically with 
$\sNN$ because diffusion increases if more energy is available -- a single exception is within the error bars. However, they are both only weakly dependent on centrality at all six energies that we have investigated. The relatively small variations  with centrality are rather unsystematic, but appear to be within the uncertainties of both data and model assumptions. For more peripheral collisions than the ones investigated here, one may expect a more systematic dependence of the diffusion coefficients on centrality as well.   The energy and centrality dependence of the particle content of the sources will be discussed in more detail in the next section.


When comparing the models for linear drift and sinh-drift in fig.\,\ref{fig1}, particle production from each fragmentation source in the linear case is almost exclusively confined to the forward- or backward-going region of pseudorapidity: The forward-going fragmentation distribution $(\eta>0)$ contains only a small fraction of particles that are produced with negative pseudorapidities $(\eta<0)$, thus moving in the opposite direction of the leading baryons, and vice versa. In the sinh-drift case, however, a substantial fraction of the fragmentation distributions spills over to the opposite-sign pseudorapidity region, corresponding to charged-hadron production in the direction that is opposite to the leading baryons in the fragmentation sources. 

Physically, this is quite possible because charged hadrons are to a large extent produced in pairs not only in the fireball, but also in the fragmentation sources.
Hence, the model with sinh-drift that has the correct Maxwell-J\"uttner limit for $t\rightarrow \infty$ is also consistent with the physical expectation for particle production in the fragmentation sources, and more realistic than the linear model that produces fragmentation distributions that are symmetric around the pseudorapidity values where the maxima of particle production occur.  Still, one may expect that a derivation of the drift term from a consistent relativistic theory for particle production such as performed in Ref.\,\cite{hgw20} for stopping can yield a more complicated form of the drift term, such that linear and sinh-drift appear as limiting cases, with the correct physics in between.

\begin{figure*}[t!]
\begin{center}
\includegraphics[scale=0.86]{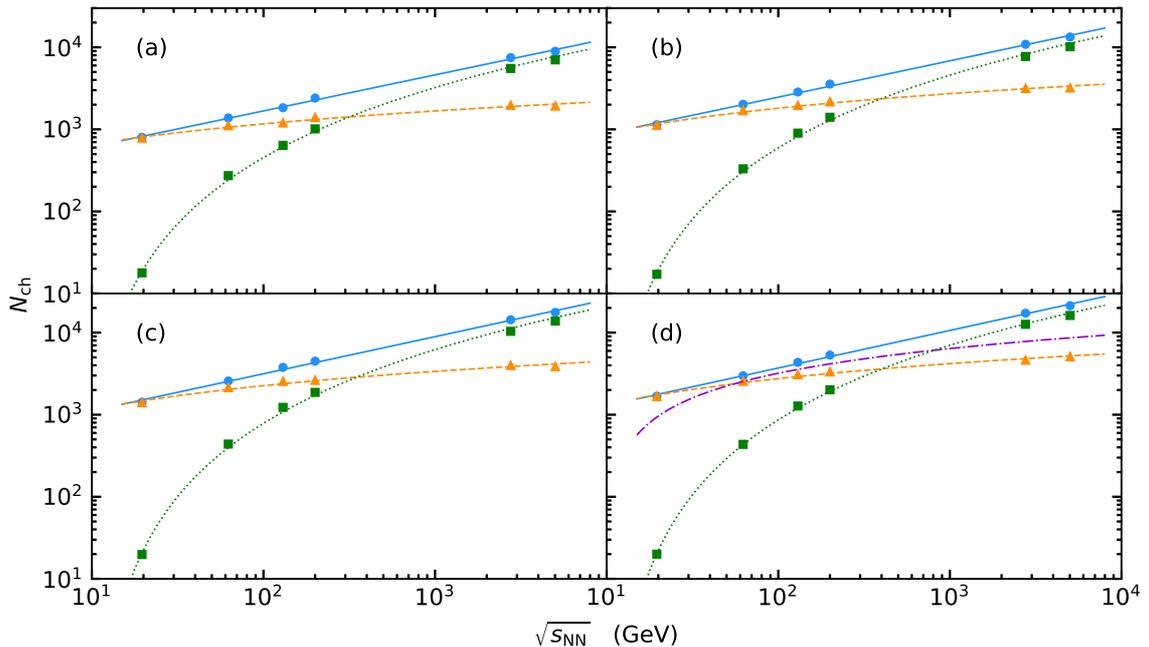}
\caption{Number of produced charged hadrons $N_\text{ch}$ as function of $\sNN$ in four different centrality classes for the three sources in the RDM with nonlinear drift. The total number of particles (circles) is fitted with a power law (solid line). The particle content of the fragmentation sources (triangles) depends logarithmically on $\sNN$ (dashed), whereas the midrapidity source (squares) shows a cubic-logarithmic (dotted) dependence. (a) Centralities $20-30$\,\% at RHIC and LHC energies. (b) $10-20$\% at RHIC and LHC. (c)  $6-10$\% at RHIC and $5-10$\% at LHC.  (d) $0-6$\,\% at RHIC and $0-5$\,\% at LHC. For comparison, the result for the fragmentation sources in the analytic model with linear drift \cite{gw15} is also shown in (d), dot-dashed.}
\label{fig3}
\end{center}
\end{figure*}

In this work, we are emphasizing the fragmentation region, where the Jacobian has almost no effect. We solve eq.\,(\ref{eq:4aa}) with sinh-drift using Dirichlet boundary conditions  in four centrality classes at six incident energies, with parameters given in table\,\ref{tab1}. We perform $\chi^2$-minimizations with respect to the data in every centrality bin using \textsc{matlab}.
The resulting charged-hadron pseudorapidity distributions are shown in fig.\,\ref{fig2}  as functions of $\tilde{\eta}=\eta-y_\text{beam}$ for $\sqrt{s_{NN}} = 19.6, 62.4, 130$ and 200 GeV Au-Au with PHOBOS data \cite{alver11} and  $\sqrt{s_{NN}} = 2.76$ and $5.02$ TeV Pb-Pb with ALICE data  \cite{abb13,alice17}. Limiting fragmentation is seen to be approximately fulfilled at all four centralities for $\tilde{y}\gtrsim 2$ as highlighted by the inset in (a) for centralities $20-30$\,\%. 

In fig.\,\ref{fig2x} we show a ratio plot of model results to data for central collisions as in case (d) of fig.\,\ref{fig2}. LF is approximately fulfilled when compared to RHIC data: The ratio ${(dN/d\eta)_\mathrm{RDM}}/{(dN/d\eta)_\mathrm{exp}}$ is reasonably close to unity in the range $\tilde{\eta}\lesssim 0$. LHC data presently terminate at $\tilde{y}\gtrsim -2.6$ for $\sqrt{s_{NN}} = 2.76$ TeV and at $\tilde{y}\gtrsim -3.7$ for $\sqrt{s_{NN}} = 5.02$ TeV. 
For data points in the range $\tilde{\eta}>0$ the ratio approaches zero rapidly because $(dN/d\eta)_\mathrm{RDM}$ drops towards zero faster than the data to fulfil the Dirichlet boundary condition imposed by the numerical solver of the FPE. Comparable results are obtained at the other three centralities investigated here.

Due to the sinh-drift, the fragmentation distributions are less confined to the fragmentation region as in the linear-drift model, but extend over the whole pseudorapidity range.
Hence, the Jacobian deforms also the fragmentation distributions in the midrapidity region, as shown already in fig.\,\ref{fig1} for central collisions.
Limiting fragmentation is clearly fulfilled within the RDM with sinh-drift  in all four centrality classes when comparing with the Au-Au data at RHIC energies, and the results are consistent with LF for Pb-Pb data at LHC energies 2.76 and 5.02 TeV in all four centrality classes as well.

As discussed before in case of central Pb-Pb at 2.76 TeV (fig.\,\ref{fig1}), the values of the diffusion strengths $\gamma_{k}$ that are needed to fit the energy- and centrality-dependent data are consistently larger than what is calculated from the relation $\gamma_k=T/m^k_\text{T}$ in the sinh-drift model, because collective expansion broadens the distributions in the three sources.
This effect increases monotonically with energy. It is also of interest to compare the deduced values of the diffusion strength in the fragmentation sources with the ones that we had obtained for stopping in central collisions of Au-Au at $\sqrt{s_{NN}}=200$ GeV as $\gamma_{1,2}=33$ in Ref.\,\cite{kgw19}. This value is significantly smaller than what we obtain in the corresponding particle-production case, $\gamma_{1,2}\simeq 61$ (see table\,\ref{tab1}), due to the fact that the fragmentation peaks in stopping are always closer to the beam rapidity than the corresponding ones in particle production. This is expressed by the larger values of $\gamma_{1,2}$, or smaller rapidity relaxation times in the fragmentation sources.

Regarding limiting fragmentation, there has been an experimental study of inclusive photons by ALICE in pp-collisions at $\sqrt{s_{NN}} = 0.9, 2.76$, and 7 TeV in the forward-rapidity region \cite{abelev15} at $\tilde{\eta}=\eta-y_\text{beam} \simeq -6.5$ to $-3$ that indicated a violation of LF in that region. Since most photons come from soft $\eta$-meson decays, hadrons would be expected to break LF if photons do. However, the pseudorapidity-region that is tested there is still too far away from the beam
rapidities ($y_\text{beam} = 6.86, 7.99$, and 8.97 at $\sqrt{s} = 0.9, 2.76$ and 7 TeV).
LF can not be expected to occur in this region, at LHC energies one must
be closer to the beam. Indeed, ALICE concludes
that at the LHC, limiting fragmentation ``may be confined to a pseudorapidty
interval closer to the beam rapidity''. In our model, it occurs at 
$\tilde{\eta}=\eta-y_\text{beam} \gtrsim  -2.5$, which has no overlap in pseudorapidity with the above ALICE results.


\begin{table*}[h]
\caption{Number of produced charged hadrons in total $N_\text{tot}$,  in the fragmentation sources $N_\text{f}$ and in the midrapidity source $N_\text{gg}$ in four centrality classes as a function of the center of mass energy squared, $s_{NN}$.}
\centering
\begin{tabular}{cccc}
\hline\hline

\multicolumn{4}{c}{$N_\mathrm{tot} (s) = a\, (s_{NN} / \SI{1}{TeV^2})^b$} \\
\hline
 
\multicolumn{1}{c}{\parbox{4cm}{\vspace{1mm}\centering Centrality class \\ RHIC / LHC\vspace{1mm}}} & 
\multicolumn{1}{c}{$a \pm \Delta a$} & 
\multicolumn{1}{c}{$b \pm \Delta b$} &
\multicolumn{1}{c}{$\chi^2/\mathrm{ndf}$} \\
\hline
0--6\,\% / 0--5\,\% & $(10.6 \pm 0.2)\times 10^3$ & $0.229 \pm 0.004$ & 17.3 \\

6--10\,\% / 5--10\,\% & $(8.9 \pm 0.2)\times 10^3$ & $0.226 \pm 0.005$ & 17.5 \\
10--20\,\% & $(6.8 \pm 0.2)\times 10^3$ & $0.221 \pm 0.005$ & 12.1 \\

20--30\,\% & $(4.6 \pm 0.1)\times 10^3$ & $0.220 \pm 0.005$ & 9.9 \\
\hline\hline

\multicolumn{4}{c}{\quad $N_\mathrm{f} = \tilde{a} \ln (s_{NN}/\tilde{s_0})$} \\
\hline
 
\multicolumn{1}{c}{\parbox{4cm}{\vspace{1mm}\centering Centrality class \\ RHIC / LHC\vspace{1mm}}} & 
\multicolumn{1}{c}{$ \tilde{a} \pm \Delta  \tilde{a}$} &
\multicolumn{1}{c}{$\tilde{s_0} \pm \Delta \tilde{s_0}$ (GeV$^2$)} &
\multicolumn{1}{c}{$\chi^2/\mathrm{ndf}$} \\
\hline
0--6\,\% / 0--5\,\% & $313 \pm 17$ & $1.5 \pm 0.7$ & 7.6\\

6--10\,\% / 5--10\,\% & $242 \pm 18$  & $0.9 \pm 0.6$ & 11.3\\
10--20\,\% & $198 \pm 10$ & $1.1 \pm 0.5$ & 4.2 \\

20--30\,\% & $111 \pm 7$ & $0.3 \pm 0.2$ & 4.0 \\
\hline\hline

\multicolumn{4}{c}{\quad $N_\mathrm{gg} =\hat{a} \ln^3 (s_{NN} / \hat{s_0})$} \\
\hline
 
\multicolumn{1}{c}{\parbox{4cm}{\vspace{1mm}\centering Centrality class \\ RHIC / LHC\vspace{1mm}}} & 
\multicolumn{1}{c}{$\hat{a} \pm \Delta \hat{a}$} &
\multicolumn{1}{c}{$\hat{s_0} \pm \Delta\hat{s_0}$ (GeV$^2$)} &
\multicolumn{1}{c}{$\chi^2/\mathrm{ndf}$} \\
\hline
0--6\,\% / 0--5\,\% & $9.1 \pm 0.3$ & $104 \pm 3$ & 23.0\\

6--10\,\% / 5--10\,\% & $7.9 \pm 0.5$  & $96 \pm 6$ & 65.5\\
10--20\,\% & $5.6 \pm 0.3$ & $88 \pm 5$ & 38.7 \\

20--30\,\% & $3.7 \pm 0.2$ & $69 \pm 5$ & 21.6 \\
\hline\hline
\end{tabular}

\label{tab2}

\end{table*}

\section{Charged-hadron content of the sources}
\label{sec4}

As proposed in Ref.\,\cite{gw15} for the RDM with linear drift, we now investigate the energy dependence of charged-hadron production in the three sources using the model with sinh-drift
for each centrality class. Since the distributions resulting from the fragmentation sources have a different shape in the nonlinear-drift model, we can expect different results from the linear-drift case.

The total number of charged hadrons that are produced in the three sources follows a power law\,\cite{gw15} 
\begin{align}
N_\text{ch}=\sum_i N_i \sim s^b\,.
\end{align} 
The variable $s$ is a dimensionless squared energy ratio $s\equiv s_{NN}/s_0$, with suitably chosen $s_0$.
The particle content produced by the fragmentation sources, $N_{1,2}$ depends logarithmically on $s$,
\begin{align}
N_{1,2} \sim \ln (s).
\end{align}  
The midrapidity source $N_\mathrm{gg}$, however, behaves differently \cite{gw15}. Its width $\Gamma$
is related to the beam rapidity $y_\text{beam}$,
\begin{align}
\Gamma \sim y_\text{beam} \simeq \ln \frac{\sNN}{m_\text{p}}= \ln(s)/2+\text{const.}
\end{align}
Hence, it is a linear function of the logarithm of $s$. The particles produced in the midrapidity source result mainly from low-$x$ gluon-gluon interactions, with $x$ the partonic longitudinal momentum fraction. The predicted cross section for such events is proportional to $\ln(s)^2$\,\cite{igi09}, satisfying the Froissart limit\,\cite{fro61}. Since the cross section is directly proportional to the yield density,
 the midrapidity source distribution scales with $\ln (s)$ in width and with $\ln(s)^2$ in yield density, such that the total functional dependence of the produced charged particles
in the central source is expected to be approximately \,\cite{gw15,gw20}
\begin{align}
N_\mathrm{gg} \sim \ln(s)^3.
\end{align}
We extract the values of $N_{1,2,\mathrm{gg}}$ from our RDM-analysis with nonlinear drift using
 suitable fit functions 
for the total number of produced charged particles,
$N_\mathrm{tot} = \sum_i N_i$,
the number of particles produced in the fragmentation sources,
$N_\mathrm{f} = N_1 + N_2$,
and the number of particles produced in the midrapidity source, $N_\mathrm{gg}$, as functions of the squared center of mass energy $s_{NN}$. The fits are done in Python using the \texttt{curve\_{}fit}-function from the \texttt{scipy.optimize} package. The routine computes the best results via a $\chi^2$-minimization and also computes the covariance matrix, where the square roots of its diagonal entries are the standard deviations.

For the total number of charged particles a power law is expected. To be able to compare our findings directly with the results for linear drift in Ref.\,\cite{gw15}, we choose $s_0=\SI{1}{TeV^2}$, such that $s = s_{NN}/ 1\,\mathrm{TeV}^2$  and
\begin{align}
N_\mathrm{tot} (s_{NN}) = a\,\left( \frac{s_{NN}}{\SI{1}{TeV^2}} \right)^b.
\end{align} 
For the most central class (see table~\ref{tab2}), we find
\begin{align*}
a &= (10.6 \pm 0.2) \times 10^3\,, \\
b &= 0.229 \pm 0.004\,,
\end{align*}
consistent with the values $a=1.1\times 10^4$ and $b=0.23$ of Ref.\,\cite{gw15} for linear drift. This is as expected, because the influence of the drift term on the total number of particles is negligibly small. 

The produced charged particles resulting from both fragmentation sources $N_\mathrm{f}$ are 
\begin{align}
N_\mathrm{f} = N_1 + N_2\,,
\end{align}
with the results  $N_{1,2}$ from section \ref{sec3} for the individual sources. For the fragmentation sources we expect a logarithmic dependence
\begin{align}
N_\mathrm{f} (s_{NN}) =\tilde{a}\ln \left( \frac{s_{NN}}{\tilde{s_0}} \right)\,.
\end{align}
Here we use $\tilde{a}$ and $\tilde{s_0}$ as parameters. Since $\ln(s_{NN}/\tilde{s_0}) = \ln(s_{NN}) - \ln(\tilde{s_0}), \tilde{s_0}$ determines the offset and should therefore be evaluated by the routine and not be fixed. This yields 
\begin{align*}
\tilde{a} &= 313 \pm 17, \\
\tilde{s_0} &=  (1.5 \pm 0.7)\,\mathrm{GeV}^2
\end{align*}
for the most central class.  The values differ from the ones in the RDM with linear drift \cite{gw15}, because the latter tends to overestimate the effect of the fragmentation sources.
In the sinh-drift model, drift and diffusion are so strong, that the fragmentation distributions spread over the whole pseudorapidity range. Hence, the form of the total distribution results in the case of a sinh-drift mainly from the central source -- but the fragmentation sources are still relevant to decide whether LF is fulfilled.

For the midrapidity source, the functional dependence on $s$ is a cubic logarithm as discussed  above,
\begin{align}
N_\mathrm{gg} (s_{NN}) =\hat{ a} \ln^3 \left( \frac{s_{NN}}{\hat{s_0}} \right).
\end{align}
The results in the most central class are 
\begin{align*}
\hat{a} &= 9.1 \pm 0.3, \\
\hat{s_0} &=  (104 \pm 3)\,\mathrm{GeV}^2.
\end{align*}
In Ref.\,\cite{gw15} the RDM with linear drift resulted in $\hat{a}=7.5$, $\hat{s_0}=\SI{169}{GeV^2}$, corresponding to a smaller yield in the central source. The difference is due to the overestimate of the fragmentation sources in the model with linear drift, which causes an underestimate of the midrapidity source.

The results for the particle numbers in the centrality classes that we have investigated in the model with sinh-drift are shown in fig.\,\ref{fig3}. In central collisions, the results for the fragmentation sources using the model with linear drift as in Ref.\,\cite{gw15} are also displayed, dot-dashed curve in (d). For the other centralities, analyses in the model with linear drift are not available. The midrapidity source becomes the main source for particle production at energies beyond the highest RHIC energy of 200 GeV, although at LHC energies the fragmentation sources still contribute substantially to charged-particle production. 
\section{Conclusion}
We have investigated the centrality dependence of the limiting-fragmentation conjecture in charged-hadron pseudorapidity distributions in Au-Au and Pb-Pb collisions at RHIC and LHC energies. A three-source relativistic diffusion model with sinh-drift, which ensures the 
correct Maxwell-J\"uttner equilibrium distribution, is the main basis of this study. It requires numerical solutions of the transport equation. For central collisions, a linear drift that allows for analytical solutions has also been tested. 

The linear-drift model produces fragmentation distributions that are symmetric around the pseudorapidity values where the maxima of particle production occur and remain essentially confined to the rapidity region of the leading baryons. In contrast, the nonlinear-drift model causes asymmetric distributions and significant fragmentation-source particle production with pseudorapidities in the opposite direction that causes a mixing of target- and projectile-fragment contributions in particle production.  This is in line with the expectation that most particles are produced in pairs not only in the fireball, but also in the fragmentation sources. Hence, the nonlinear-drift model physically appears to be more realistic.

The numbers of produced charged hadrons in the three sources as functions of incident energy have been deduced. The particle number in the midrapidity source depends cubic-logarithmically on $\sqrt{s_{NN}}$ and becomes the main source of particle production at LHC energies, but the fragmentation sources remain relevant and are essential to maintain limiting fragmentation.

Our analysis shows that the RDM with three sources displays approximate limiting-fragmentation scaling in agreement with the data in four centrality classes at RHIC energies. This is essentially due to the diffusion in the fragmentation sources, it would not occur with the fireball source alone.
According to our results, limiting fragmentation is also likely to be valid in the corresponding centrality classes at LHC energies, thus spanning a factor of almost 260 in collision energy. 

This conclusion disagrees with expectations from simple parametrizations of the rapidity distributions, and also with predictions from the thermal model, which does not consider the fragmentation sources. The latter play an essential role in our nonequilibrium-statistical approach.
It would be desirable  that future upgrades of the LHC detectors will make it possible to test the limiting-fragmentation conjecture experimentally at LHC energies in different centrality classes.

\begin{acknowledgement}
\end{acknowledgement}
One of the authors thanks Johannes H\"olck (ITP Heidelberg) and Klaus Reygers (PI Heidelberg) for discussions, in particular, about the form of the drift term in the RDM.
We are grateful to the referees for constructive remarks.
\bibliographystyle{epj}
\bibliography{kgw20.bib}
\end{document}